\documentclass[%
reprint,
superscriptaddress,
amsmath,amssymb,aps, prb, floatfix, showpacs,citeautoscript
]{revtex4-1}

\usepackage[]{graphicx}
\usepackage{dcolumn}
\usepackage{bm}
\usepackage{amsfonts,amssymb,amsmath}
\usepackage[colorlinks,linkcolor={blue},citecolor={blue},urlcolor={blue}]{hyperref}
\usepackage{textcomp}
\usepackage{float}

\usepackage{color}
\usepackage{array}
\usepackage{booktabs}
\usepackage{multirow}
\usepackage{color}
\begin{document}

	\title{Topological phases and twisting of graphene on a dichalcogenide monolayer}
	
	\author{Abdulrhman M. Alsharari}
	\email{aalsharari@ut.edu.sa}
	\affiliation{Department of Physics and Astronomy, and Nanoscale and Quantum
		Phenomena Institute, \\ Ohio University, Athens, Ohio 45701, US}
	\affiliation{Department of Physics, University of Tabuk, Tabuk 71491, Saudi Arabia}
	\author{Mahmoud M. Asmar}
	\affiliation{Department of Physics and Astronomy, Louisiana State University, Baton Rouge, Louisiana 70803, US}
	\author{Sergio E. Ulloa}
	\affiliation{Department of Physics and Astronomy, and Nanoscale and Quantum
		Phenomena Institute, \\ Ohio University, Athens, Ohio 45701, US}

	\date{\today}

	\begin{abstract}	
		Depositing monolayer graphene on a transition metal dichalcogenide (TMD) semiconductor substrate has been shown to change the dynamics of the electronic states in 
		graphene, inducing spin orbit coupling (SOC) and staggered potential effects.  Theoretical studies on commensurate supercells have demonstrated the appearance of interesting phases, as different materials and relative gate voltages are applied.  Here we address the effects of
		the real incommensurability between lattices by implementing a continuum model approach that does not require small-period supercells.  The approach allows us to study the role 
		of possible relative twists of the layers, and verify that the SOC transfer is robust to twists, in agreement with observations.  
		We characterize the nature of the different phases by studying an effective Hamiltonian that fully describes the graphene-TMD heterostructure.  We find
		the system supports topologically non-trivial phases over a wide range of parameter ranges, which require the dominance of the intrinsic SOC over the staggered and Rashba
		potentials.  This tantalizing result suggests the possible experimental realization of a tunable quantum spin Hall
		phase under suitable conditions.  We estimate that most TMDs used to date likely result in weak intrinsic SOC that would not drive the heterostructure into 
		topologically non-trivial phases.  Additional means to induce a larger intrinsic SOC, such as strain fields or heavy metal intercalation may be required. 
	\end{abstract} 
	
	\maketitle

\section{Introduction}
		
	Depositing graphene \cite{graphene,graphene2} (G) on transition metal dichalcogenide \cite{3B-TB-TMD,mos2,mos22} (TMD) 2D layered materials has attracted much recent attention as a novel way of tailoring graphene properties \cite{Inv-G-TMDs,mogrf,mogrf2,ferreira}.  
	The successful experimental growth of G-TMD heterostructures has provided intriguing results  \cite{Band-Alignment2,G-MOS2-exp,SOC-in-graphene,w2,Band-Alignment1,SOC-G-TMDs,G-WS2-Exciton,G-WS2-Edge_SOC,Wakamura}. 
	For instance, G-WS$_2 $ heterostructures have been shown to result in the enhancement of spin-orbit coupling (SOC) in graphene without appreciable effect on its characteristic high mobility \cite{SOC-in-graphene}, and the asymmetry of a monolayer appears to produce stronger SOC than a bulk substrate \cite{Wakamura}.
	 However, this heterostructure is predicted to have a gap with mass-inverted bands and interesting spin structure \cite{w2}, which would result in unusual transport properties \cite{mahmoud-PRB}, including the appearance of chiral conducting channels on the structure edge \cite{mogrf2,Frank-GT}.  Edge channels
may give rise to the appearance of a quantum spin Hall effect (QSHE) with obvious spintronic applications, as counterpropagating channels with opposite spin would carry spin currents without dissipation \cite{qshe}.
	Other experiments report a gapless band structure at the neutrality point in epitaxial G-MoS$_2$ heterostructures, although exhibiting several miniband-gaps at larger energies \cite{Band-Alignment1}. These features are attributed to an orientational twist in the graphene layer with respect to the TMD, estimated to be $\simeq 6^\circ$, which would create moir\'e patterns as those seen on G-hBN \cite{Ele.Prop.G-hBN}.
	
	The lattice mismatch between graphene ($a=2.46$\AA) and TMDs (e.g., $\tilde{a}=3.11$\AA{ } for MoS$_2$) has led to the study of various supercells with approximately commensurate structures. These have been studied by means of density functional theory \cite{w2,mogrf,mogrf2,Sobhit} (DFT) and tight-binding formalisms \cite{Inv-G-TMDs}. Most investigated supercells utilize size ratios of 9/7 (9 graphene cells to 7 TMD cells), 5/4, and 4/3 \cite{w2}. Results for these supercells show similarities, with quantitative differences arising from strains generated on the constituent layers at the different size ratios. For example, structures of 9/7 and 5/4 supercells of G-WS$_2$ show inverted-mass bandgaps in DFT calculations \cite{G-WS2-Edge_SOC}. 
A 4/3 structure, however, displays direct or inverted-mass bandgaps, depending on the choice of TMD substrate \cite{mogrf,mogrf2}. 
This structure may also show slight charge transfer to the TMD, likely the result of the built-in strains \cite{Sobhit}.
Other 5/4 structures show inverted-mass bands, as well as a phase transition upon the application of a gate voltage between the two layers of the system \cite{Inv-G-TMDs}. 

Given the natural incommensurability between G and TMD layers, it is natural to ask if such mass-inversion and transitions also exist in a general misaligned heterolayer system.  It is also interesting to see how the effective SOC in graphene due to proximity to the TMD evolves with relative interlayer twisting.	
This work studies misaligned G-TMD layers, as the influence of incommensurate lattices on the electronic structure has not been fully addressed. Experimentally,
studies of  mono and multilayered graphene deposited on various TMD substrates with different orientational alignments have been reported \cite{SOC-G-TMDs,Band-Alignment1}. 
They find that SOC is robust against the number of graphene layers, TMD material used as a substrate, and different rotational misalignments. We undertake a theoretical analysis of this problem utilizing a continuum model approach \cite{G-Bi-w-Twist-Continuum.Model,G-Bi-w-Twist-.Low.Energy,Moiré-bands-in-twisted-double-layer-graphene}. It captures the hybridization between the graphene and the TMD layers correctly for both commensurate as well as incommensurate systems. This allows us to explore, in particular, how the electronic structure evolves as the TMD substrate rotates with respect to the graphene lattice.
	
	Once the continuum model structure is solved, the results are parameterized using an effective Hamiltonian that includes all relevant symmetry-allowed terms.   
Such Hamiltonian has been proven to capture all the essential features of the different systems, in agreement with various microscopic calculations,
and allows us to extract the SOC and other relevant parameters, as well as their dependence on relative layer rotation and other structural features. 
Results reveal, for example, a pronounced SOC for all rotation angles, in agreement with recent experiments \cite{SOC-G-TMDs}.
Moreover, we find an overall drop in the staggered potential field as the twisting angle increases, as one would anticipate.  The presence of both SOC and staggered interactions results in a competition that can produce mass inversion as function of angle and/or applied gate voltage in a given structure.  Such readily available physical perturbations allow one to explore the
phase diagram of these interesting structures, augmenting our ability to tune the properties of a system aS will.	
	
 We also show that adequate manipulation of the system drives it to exhibit multiple interesting phases. Some of these have been identified and studied numerically using calculations of the $\mathbb{Z}_2$ invariant \cite{Frank-GT}, as well as  edge state identification in finite size systems.
	In this work, we present a systematic exploration of the parameter ranges of the effective model, in order to identify all possible topological regimes, study their interesting features, and find analytical conditions to describe gapless phases and identify topologically distinct regimes. We determine the topological features of different phases based on $\mathbb{Z}_2$ invariant characterization, and validate the analytical conditions that govern the existence of non-trivial regimes.
	
The model Hamiltonian allows for a variety of interesting topologically non-trivial phases for moderate values of different interaction terms.  This suggests that it may be possible
to achieve QSHE phases in heterolayer systems for the appropriate material and applied gate regimes.  However, we
also find that realistic microscopic parameters that describe the most frequently used TMDs may produce only topologically trivial phases in experiments. 
Reaching non-trivial topologies will likely require further enhancement of SOC, so as to make it the dominant interaction.  This may be obtained by applied electric fields and/or intercalation of heavy transition metals \cite{Berlin}, for example. 

In what follows, the continuum model to treat incommensurate lattices, as well as typical results are presented in section \ref{model}.  That section also discusses the effective Hamiltonian that
describes the near-gap behavior of the G-TMD system.  In particular, section \ref{rotation} shows how relative interlayer twists may give rise to a gap closing regime separating direct and inverted-mass bandgaps, by monitoring the twist-dependence of the different effective Hamiltonian parameters.  The following two sections develop a more rigorous characterization of the different phases
of the Hamiltonian by considering when a semimetallic regime is reached (gap closing, section \ref{closing}), and what is the topological index (or $\mathbb{Z}_2$ invariant) for such phases (section \ref{Z2}).  In the last section we provide a brief discussion emphasizing the relevance of these results.

\section{Continuum Model} \label{model}
	In order to study the effect of the lattice incommensurability or relative rotation of a graphene layer placed in proximity to the TMD, we build a continuum model to describe such general  heterostructure \cite{G-Bi-w-Twist-Continuum.Model,G-Bi-Twist-Band-Symmetry-singularity,G-Bi-Twist-Comm-Coherence,G-Bi-Twist-Ele.Str.,G-Bi-Twist-Quan.Interf.,G-Bi-w-Twist-.Low.Energy,Moiré-bands-in-twisted-double-layer-graphene}. Although details of our calculations are given in the supplement \cite{supp}, we provide here a brief description. The continuum model Hamiltonian is described in terms of the independent constituent subsystems and an interlayer coupling term, so that 
$\mathcal{H}_{c}=\mathcal{H}_{g}+\mathcal{H}_{t}+\mathcal{H}_{u}$. $\mathcal{H}_{g}$ 
describes the graphene effective Hamiltonian with linear dispersion near two inequivalent valleys in its hexagonal Brillouin zone, and $\mathcal{H}_{t}$ is the corresponding Hamiltonian for the TMD
monolayer that describes the electronic dispersion in a three-orbital representation \cite{supp}.  The interlayer coupling is written as,
	\begin{equation}\label{coupling}
	\begin{split}
	\mathcal{H}_{u}=-  \sum_{\boldsymbol{G}\boldsymbol{\tilde{G}}}e^{-i\boldsymbol{G}\cdot\boldsymbol{\tau}_{X}+i\boldsymbol{\tilde{G}}\cdot\boldsymbol{\tau}_{\tilde{X}}}  t_{X{\tilde{X}}}(\boldsymbol{k}+\boldsymbol{G}) \delta_{\boldsymbol{k+G,\tilde{k}+\tilde{G}}} \, , 
	\end{split} 
	\end{equation} 
where $\boldsymbol{\tau}_{X}$ and $\boldsymbol{\tau}_{\tilde{X}}$ are the sublattice positions in the unit cells of each of the layers.  $\boldsymbol{G}$ and  $\boldsymbol{\tilde{G}}$ are reciprocal lattice vectors that solve the in-plane momentum conservation condition in the interlayer tunneling process,  
$\boldsymbol{k}+\boldsymbol{G}=\boldsymbol{\tilde{k}}+\boldsymbol{\tilde{G}}$.
The interlayer transfer integral between graphene $p_z$ and TMD $d$-orbitals, $X$ and $\tilde{X}$, respectively, is $t_{X{\tilde{X}}}(\boldsymbol{q})$ \cite{slater}. 
	This integral depends on the momentum $\boldsymbol{q}$, measured from the $\Gamma$ point, and falls quickly to zero as $\boldsymbol{q}$ increases.
We note that since the incommensurability is small, the hybridization between  K and K$ ' $ valleys is negligible at low energy, as the large momentum separation protects against intervalley scattering \cite{incomen}.	Hence, each valley is reliably treated separately, as long as the relative twist is small, e.g., $\theta \lesssim 5^\circ$ \cite{Moiré-bands-in-twisted-double-layer-graphene,Ele.Prop.G-hBN,mori3}.
	It should be noted that the existence of misaligned layers and finite $q$-range of the interlayer coupling requires the inclusion of multiple K valleys that solve the  momentum conservation condition. The sum in Eq.\ \eqref{coupling} is easily continued to achieve convergence to the desired accuracy.

\subsection{G-TMD Hamiltonian at zero relative twist} \label{notwist}
	
\begin{figure}[t]
		\centering
		\includegraphics[width=\linewidth]{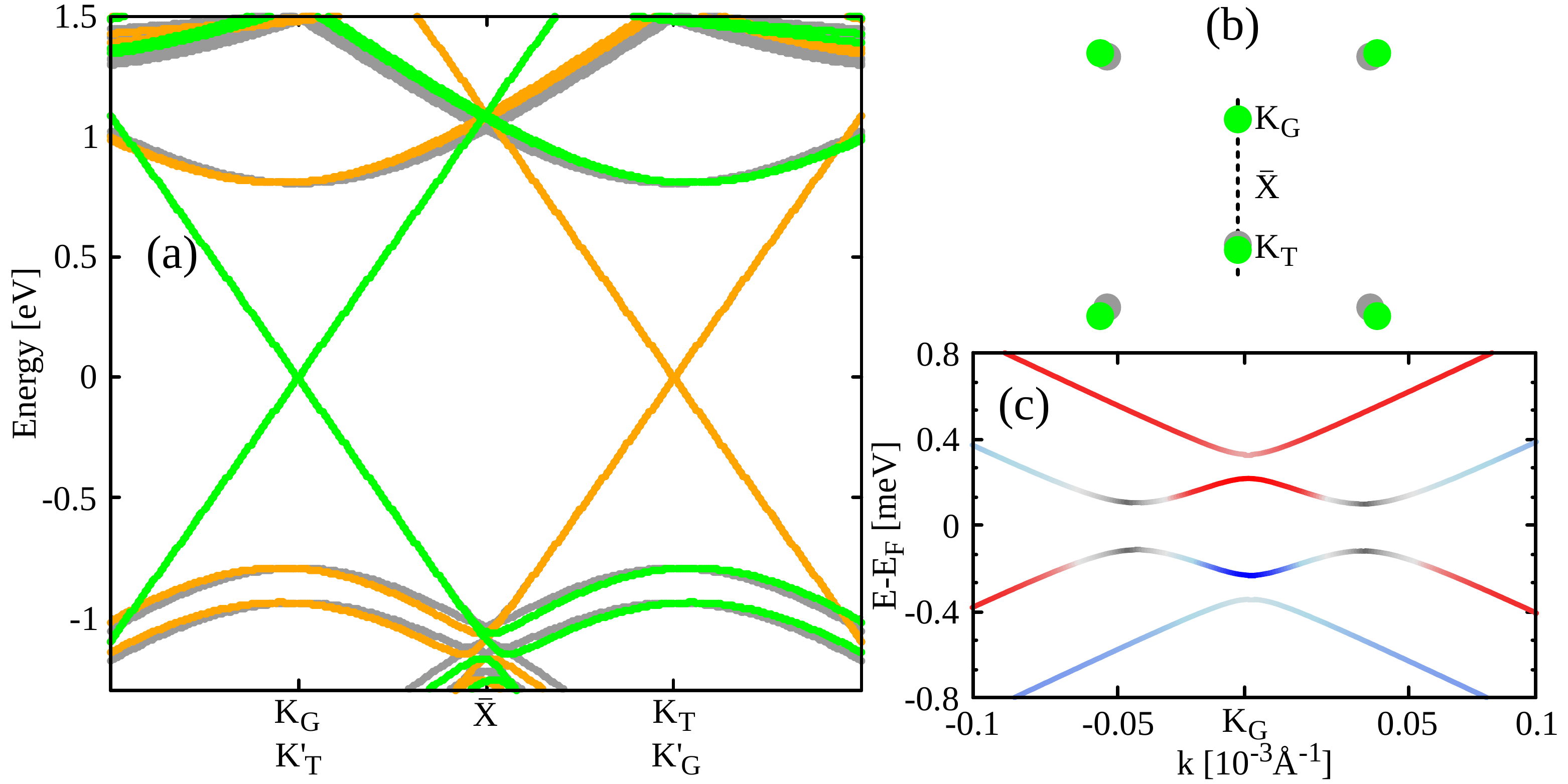}
		\caption{(Color online) (a) Band structure of coupled graphene-TMD continuum model at zero relative rotation angle, shown along the reciprocal space direction indicated in panel (b).  The commensurate continuum model is shown in gray lines; the incommensurate model results are shown  in green and orange lines.
			The small incommensurability in real space makes the TMD K$_{\rm T}$ point (orange lines) to fold close to the K$_{\rm G}$ graphene point  (green lines).
Panel (b) shows the relative positions of nearby points (green) in comparison to those for the commensurate structure (gray) around the  K$_{\rm G}$ point of interest.
			(c) Zoom-in into the graphene neutrality points reveals band splitting due to Zeeman-like and Rashba SOC effects, leading here to mass inverted bands. Due to time reversal symmetry, K$'_{\rm G}$ valley shows similar features with spin reversed states (not shown).}
		\label{Fig3}
	\end{figure}

	Figure \ref{Fig3} shows the band structure of a commensurate (gray bands, 5/4 structure) and incommensurate (colored bands) continuum  model along the K$_{\rm G}$-K$_{\rm T}$ path shown in panel (b). 
	 Despite the natural incommensurability in the system, the low energy results of the two models agree well and show similar structure.
	Our analysis shows that both continuum model descriptions of the G-TMD at zero angle exhibit similar main characteristic features, such as direct and inverted band gaps, and are also in agreement with tight-binding calculations of commensurate supercells \cite{Inv-G-TMDs}.
	Some of these features include the response of the system to the application of an effective gate voltage that shifts the graphene bands with respect to those of the TMD.  The gate voltage drives a transition between phases that have a simple direct (quadratic) bandgap structure and an inverted-mass regime with more complex structure  \cite{Inv-G-TMDs}.
The latter regime/behavior is seen when the  neutrality point of graphene is driven close to the TMD valence bands, as shown in Fig.\ \ref{Fig3}c.
	We note that the continuum model agrees with the 5/4 supercell tight-binding supercell results, as discussed in the supplement \cite{supp}.

	The minimal effective Hamiltonian that describes such  G-TMD heterostructures near the neutrality point has the form \cite{Inv-G-TMDs,Frank-GT}
	\begin{equation}\label{Effective_H}
	\mathcal{H}_{\rm{eff}}=\mathcal{H}_{0}+\mathcal{H}_{\Delta}+\mathcal{H}_{S}+\mathcal{H}_{\lambda}+\mathcal{H}_R ,
	\end{equation}
	with
	\begin{equation}\label{Eff_H}
	\begin{split}
	\mathcal{H}_{0 } &= v_F  \mathit{s}_{0 }( \tau _z \sigma _x  \mathit{p}_x+ \tau _{0 } \sigma _y \mathit{p}_y  ),\\	
	\mathcal{H}_{\rm{\Delta} } &=\Delta \mathit{s}_{0 } \tau _{0 } \sigma _z ,\\
	\mathcal{H}_{\rm{S}}&= S  \mathit{s}_z \tau _z \sigma _z,\\
	\mathcal{H}_{\rm{\lambda}} &=\lambda  \mathit{s}_z \tau _z \sigma _{0 } ,\\
	\mathcal{H}_{\rm{R}} &=\tfrac{1}{2} R \left(  \mathit{s}_y \tau _z \sigma _x- \mathit{s}_x \tau _{0 } \sigma _y \right) , \\
	\end{split}
	\end{equation}
	where  we use the `standard' basis $\Psi^T = (\Psi^T_K, \Psi^T_{K'})$, with $\Psi_{K,K'}^T= (A\uparrow,B\uparrow,A\downarrow,B\downarrow)_{K,K'}$, and Pauli matrices
 describe operators in valley space, $\tau_j$, pseudospin, $\sigma_j$, and spin, $s_j$,  with $j=0,x,y\text{ and } z$, and $j=0$ describes the unit matrix.  {$\mathcal{H}_0$}
	describes pristine graphene, where $\boldsymbol{p}=(p_x,p_y)$ is the momentum measured from each K, K$'$ valley.
	
	The low energy spectrum and eigenstate structure of the continuum model are very well described by Eq.\ \eqref{Effective_H} in different regimes.  In the case of band inversion, the system shows a Zeeman-like SOC, $\lambda$, that is larger than both the staggered term, $\Delta$, and the intrinsic SOC, $S$.  In the direct band gap regime, the staggered term dominates. 
	The Rashba interaction $R$ also contributes to the gap and spin structure of the spectrum.  The different contributions will be analyzed in more detail in the next section.

\subsection{Relative twist of the G-TMD heterostructure}\label{rotation}
	We want to explore the effect of rotating the TMD layer with respect to that of graphene. A similar question has been studied for the graphene on hBN system \cite{Ele.Prop.G-hBN}. Here, the SOC transferred from the TMD layer allows for eigenstate spinor structure that depends both on the rotation angle and the position of the graphene neutrality point within the gap of the TMD.
	
\begin{figure}[t]
		\centering
		\includegraphics[width=1.05\linewidth]{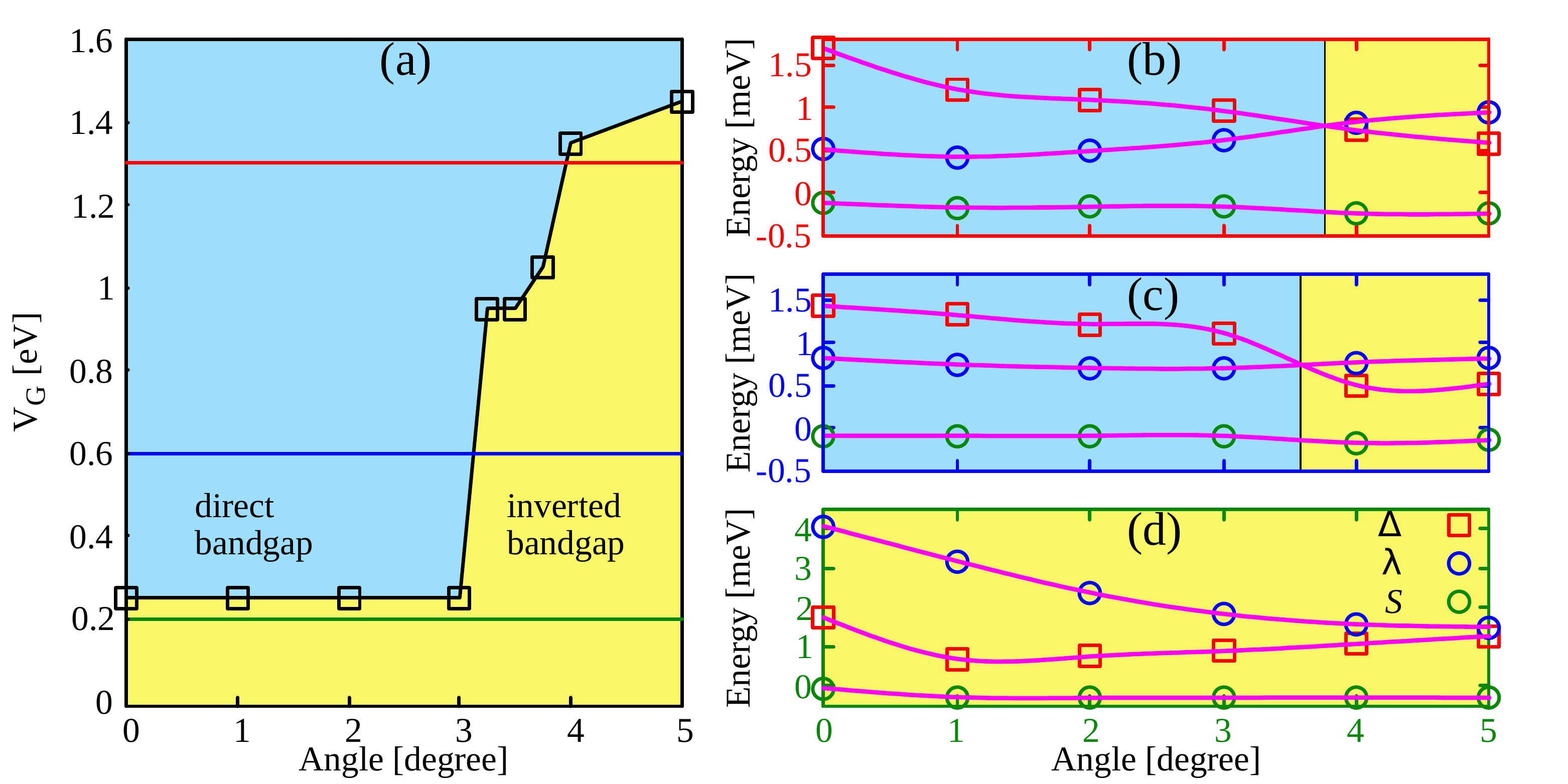}
		\caption{(Color online)  Effective parameters for rotation angles $\theta \leq 5^\circ$ in the continuum model. (a) Phase diagram on the gate voltage and twist angle plane showing the boundary between direct (blue shading) and inverted mass (yellow) bandgaps. Right panels show heterostructure effective parameters where the graphene neutrality point is close to (b) the conduction, (c) nearly at the middle, and (d) valence bands. 
			Intrinsic SOC parameter $S$ is nearly zero at all angles (green symbols). The staggered term $\Delta$ (red squares) is larger [smaller] than the Zeeman-like SOC $\lambda$ (blue circles), in (b) and (c) [(d)] panels for small twist angles. The system exhibits a direct bandgap whenever $\Delta > \lambda$, and inverted-mass bands for $\Delta < \lambda$, as
indicated by the shading. Notice that for large twists, $\theta > 4^\circ$, the system in (b) and (c) have switched to an inverted mass regime.  In all these panels, $R=0.5$ meV. Structure parameters correspond here to graphene on WS$_2$.} 
		\label{Fig4}
	\end{figure}

	Figure \ref{Fig4}b--d shows effective Hamiltonian parameters for different cases of applied gate voltage to shift the neutrality point in graphene within the gap of the TMD. 
	For zero relative twist, $\theta=0$, and neutrality point close to the conduction band, the system shows a simple direct bandgap structure, Fig.\ \ref{Fig4}b.  However, 
the system undergoes a phase transition from direct to inverted-mass bandgap as the rotation angle increases, here at $\theta \simeq 3.8^\circ$.  The relative twist makes the effective staggered potential $\Delta$ in the system decrease with angle, while the Zeeman-like SOC $\lambda$ increases slightly to overtake it. On the contrary, when the neutrality point in graphene is close to the valence bands of the TMD, Fig.\ \ref{Fig4}d shows the effective parameters as a function of twist angle. In this case, the system exhibits inverted bands for all angles in this range. 
	Although the SOC is robust, it is seen to slightly drop as a function of twist angle, which also makes the overall bandgap decrease. 
		
	This behavior suggests that one can explore different band structure regimes in the G-TMD heterolayer by not only changing the gate voltage \cite{Inv-G-TMDs}, but also the relative interlayer twist.  Figure \ref{Fig4}a shows a characteristic diagram of the direct and inverted-mass regimes as gate voltage and twist angle change.  For small gate voltage the graphene neutrality point is close to the valence bands of TMD.  In that regime, the system shows a robust inverted mass (see panel d).  At higher gate voltage the neutrality point moves towards the conduction band of the TMD, and the system changes to a direct (quadratic) band for small twist angles, but it returns to the inverted-mass regime at larger twist.  Figure \ref{Fig4}a provides the approximate phase boundary for this heterolayer system.
		
	The general robustness of the transferred SOC with either gate voltage or twist angle is in general agreement with the reported  experimental observations of strong SOC 
in G-TMD systems with generic (likely non-zero) relative rotation. For instance, Ref.\ \onlinecite{SOC-G-TMDs} finds robustness of the produced SOC against lattice misalignment and rotation faults in their G-TMD heterostructures. Interestingly, a closely related system, germanene deposited on MoS$_2$, is predicted to exhibit similar behavior, with strong induced SOC at different rotation angles \cite{Ge-MoS2}.

\section{Gap closing equations} \label{closing}
	 We have seen that the G-TMD system allows for different band structure regimes, fully described by the low energy effective Hamiltonian in Eq.\ \eqref{Effective_H}.
	 The characteristic features of this heterostructure are controlled by the bands near the two inequivalent K points, where the structure exhibits gaps with different intrinsic features. 
In general, different phases that may be inequivalent topologically would be reached in parameter space via a gapless semimetallic regime \cite{TI-SC-review,TI-Gauge-smooth}.  One is then interested in identifying the gap closing conditions at the K and K$'$ points for the effective Hamiltonian. This occurs at degeneracies of the various eigenvalues, given by these three conditions  
	 \begin{equation}\label{Cond2}
	 \begin{split}
	 \Delta+\lambda=&0,\\
	 R^2+(\Delta-\lambda)^2=&(2S\pm(\Delta + \lambda))^2 . 
	 \end{split}
	 \end{equation}
	 Notice one recovers the Kane-Mele Hamiltonian by setting $\lambda=0$ \cite{qshe}. In that case, for $2S>\Delta>0$, the gap closing separates two distinct phases: A system with $ \sqrt{R^2+\Delta^2}\leq|2S-\Delta|$ exhibits QSHE, and has a trivial topology otherwise \cite{TI-Z2-Cs}. 
	 
\begin{figure}
	 	\centering
	 	\includegraphics[width=0.99\linewidth]{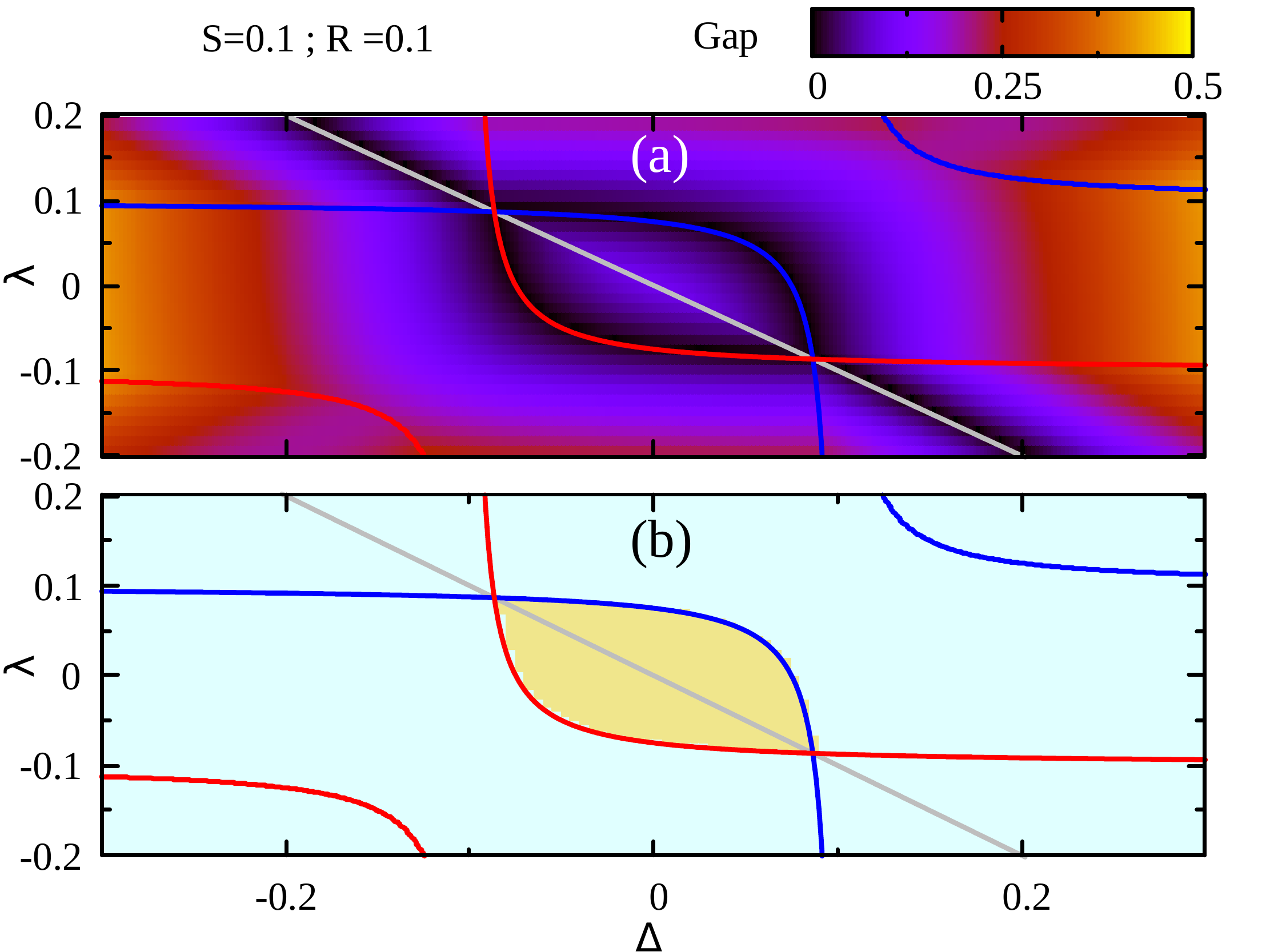} 
	 	\caption{(Color online) (a) Map of the gap size on the $\lambda$ and $\Delta$ parameter-plane for the effective Hamiltonian ($S=R=0.1$ are kept fixed). Color indicates gap size separating conduction and valence bands in the G-TMD structure. (b) $\mathbb{Z}_2$ values for the same parameter regime. Khaki (cyan) shading indicates $\mathbb{Z}_2= -1$ (+1) value,
which identifies a non-trivial (trivial) phase of the system.  Red, blue and gray curves in both panels represent band degeneracy conditions in Eq.\ \eqref{Cond2}. All effective parameters and gap size are in units of the graphene intralayer coupling, $t={2\hbar v_{\rm F} }/{a\sqrt{3}}=3.03$ eV--see supplement \cite{supp}.  }
	 	\label{Gap-Z2Phase}
	 \end{figure}

	 To validate the conditions in Eq.\ \eqref{Cond2}, we have explored different cuts of the phase diagram along different parameter planes (among $\Delta$, $S$, $R$, or $\lambda$), while keeping the other two at different constant values.  Figures \ref{Gap-Z2Phase}a and \ref{Fig44}a show that Eqs.\ \eqref{Cond2}  capture correctly the gap closings, 
and specify possible changes in the topological features of the structure. The different phases will be identified through the $\mathbb{Z}_2$ invariant we discuss in the next section.
	 
\section{$\mathbb{Z}_2$ Invariant} \label{Z2}
	 To better understand the topological characteristics of the different phases, let us first analyze the effective Hamiltonian in Eq.\ \eqref{Effective_H}.	
	 It preserves time reversal symmetry, represented by the operator $\mathcal{T}= i \tau_{0}  \sigma_x s_{y}  \mathcal{C}$ \cite{mahmoud-PRB}, where $\mathcal{C}$ is the complex conjugation operator. As such, the different phases of the system can be classified topologically through the calculation of the $\mathbb{Z}_2$ invariant number \cite{Z2-Method,Z2-Method-1,Z2-Method-2,supp}. This index distinguishes between a normal (or trivial) insulator and a non-trivial (or topological) insulator that supports edge states and the appearance of the  QSHE \cite{qshe}. 
	 
	 We have implemented a calculation of the $\mathbb{Z}_2$ invariant that follows a manifestly gauge invariant formulation of the adiabatic connection \cite{Z2-Method-2}, and has proven reliable in identifying topological phases in other systems. 
	 %
	 We find that the system in Eq.\ \eqref{Eff_H} is in a QSHE regime whenever the following conditions are met
	 	 \begin{equation}\label{condition}
	 	 | \sqrt{R^2+(\Delta-\lambda)^2} \pm(\Delta+\lambda) | < |2 S|.
	 	 \end{equation}
	 	 
	 Figure \ref{Gap-Z2Phase}b illustrates regimes in parameter space where trivial ($\mathbb{Z}_2=1$) and non-trivial  topological phases ($\mathbb{Z}_2=-1$) are separated by the conditions in Eq.\ \eqref{Cond2}. We find that the conditions in Eq.\ \eqref{condition} reliably identify the trivial and non-trivial regimes of the system, as per their corresponding $\mathbb{Z}_2$ index values.
	 We note that Frank \textit{et al.} \cite{Frank-GT} have calculated a phase diagram in the $S$ and $\lambda$ plane for fixed $\Delta$ and $R$ for this Hamiltonian, and our results are in full-agreement with their identification of phases.  
	 We note that the topological conditions in Eq.\ \eqref{condition} effectively require the intrinsic SOC $S$ to be the dominant interaction.  For $R \simeq 0$, for example, they require $|S| > |\Delta|, |\lambda|$ for non-trivial topological behavior.

\section{Discussion and Conclusions}
	The effective Hamiltonian parameters obtained from the G-TMD continuum model results prove the important role that the Zeeman-like SOC $\lambda$ plays in determining the behavior of the system. In fact, although our previous numerical estimates suggested the gap closing to occur at $\Delta+\lambda+S+R/3\simeq0$  
\cite{Inv-G-TMDs}, the analysis here shows that the gap closing in the model is given by Eq.\ \eqref{Cond2}, and mostly by the first condition ($\Delta + \lambda =0$), for the parameters studied before \cite{Inv-G-TMDs}. 
	Consideration of the $\mathbb{Z}_2$ condition in Eq.\ \eqref{condition} for these parameters confirms that these systems with direct band gap have indeed a trivial topological character ($\mathbb{Z}_2=1$). 
	Moreover, although the system can cross through a semimetallic state that separates direct and inverted-mass bandgap regimes, the $\mathbb{Z}_2$ invariant still
	identifies them as having trivial topology.  In other words, although the inverted-mass regime exhibits edge states with unique features \cite{Inv-G-TMDs,Frank-GT},  the 
	phase is {\em not} in a true QSHE state.  The edge states in the inverted-mass regime exhibit counterpropagating definite spin channels along zigzag-terminated flakes, as reported \cite{Frank-GT}. However, the preserved time reversal symmetry allows for mixing of edge states by short-range scattering defects, as well as mixing via bulk-like states \cite{Frank-GT}.   
	
\begin{figure}
		\centering
		\includegraphics[width=0.99\linewidth]{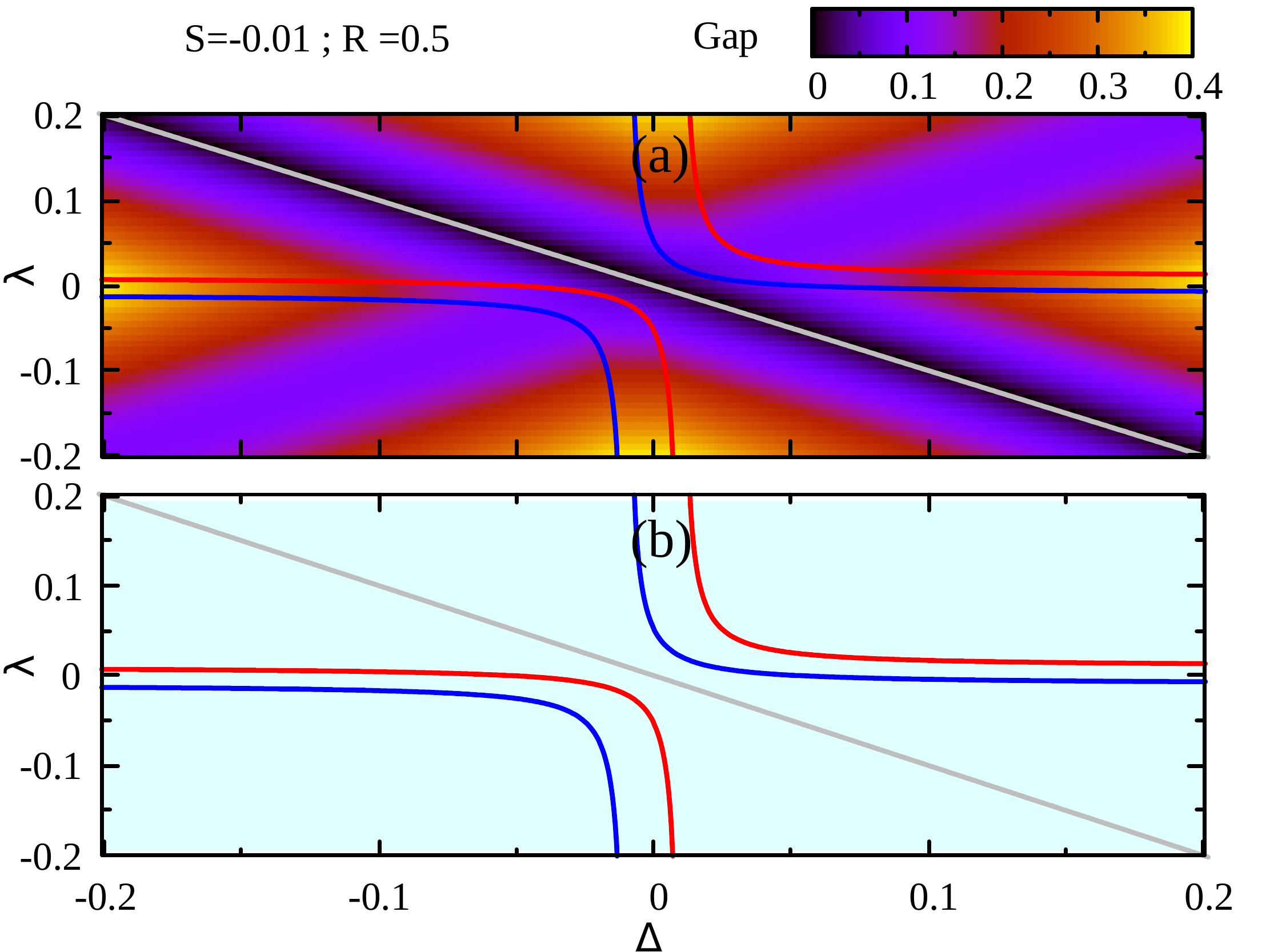} 
		\caption{(Color online) (a) Gap size map for $\lambda$ and $\Delta$ parameters of the effective Hamiltonian ($S$ and $R$ are fixed as indicated). Color represents gap size  separating conduction and valence bands in the G-TMD structure. (b) $\mathbb{Z}_2$ values for the same parameter regime. The cyan shading indicates $\mathbb{Z}_2=1$, identifying as trivial all the phases over the entire parameter regime, given the small $|S|$ and relatively large $R$ value, despite the gap closing at $\lambda = -\Delta$.  Red, blue and gray lines in both panels represent band degeneracy conditions in Eq.\ \eqref{Cond2}. All energies in units of the interlayer coupling term, $t=3.03$ eV.  }
		\label{Fig44}
	\end{figure}

	As mentioned above, the effective Hamiltonian describing the G-TMD system requires the intrinsic SOC to be the dominant interaction. To illustrate this point, Fig.\  \ref{Fig44} shows the gap map and $\mathbb{Z}_2$ phase diagram of a system where the intrinsic SOC $S$ is small. In this regime, all phases are topologically trivial regardless of the band gap morphology and associated Berry curvature, which includes here direct and inverted-mass bands \cite{supp}.  
	
	As seen in Fig.\ \ref{Fig4} and in the supplement \cite{supp}, reasonable estimates of Hamiltonian parameters for typical G-TMD structures in experiments result in
relatively large Zeeman-like SOC and staggered potentials, which are then the dominant parameters induced onto the graphene layer, with correspondingly weaker intrinsic SOC $S$. However, as the relative interlayer twist reduces the effective staggered potential, it is suggestive to employ large twists and/or higher incommensurability systems in order to reduce that effect.
	
	We have also discussed the topological invariant number $\mathbb{Z}_2$, which shows that G-TMD heterostructures are likely to be topologically trivial, despite gap  closing transitions that signal the appearance of inverted-mass bands.  
	Notice that the dominance of Zeeman-like SOC leads to an opposite shift in spin polarized bands, causing bandgap closing, although without mixing.  The presence 
of a Rashba perturbation, however, opens gaps around the K  and K$ ' $ points.  
	From the condition in Eq.\ \eqref{condition}, we expect a weak Rashba SOC to play a minimal role and not to affect the topological character of the system.  
	
	The crucial factors that determine the topological phase of the Hamiltonian in Eq.\ \eqref{Effective_H} can be seen if we first neglect the Rashba SOC term. In that case, the Hamiltonian preserves $ S_z $ and time reversal, so that the spin up and down blocks of the Hamiltonian are related by $ H_{\uparrow}(k)=-H_{\downarrow}(-k)$. Thus, subtraction of the Chern numbers for each block yields $C_{\uparrow}-C_{\downarrow}=2C_{spin}$, where $C_{spin}$ is the spin Chern number \cite{Inv-G-TMDs,Spin_Chern}. %
	This quantized invariant is equivalent to  the $\mathbb{Z}_2$ topological index, $\mathbb{Z}_2=C_{spin} \mod 2$, as long as the spin symmetry is preserved. 
	For the spin preserved Hamiltonian, it is readily found that $C_{\uparrow(\downarrow)}$ depends only on $\Delta$ and $S$, whereas $\lambda$ plays no role. 
	Thus, the dominance of $\lambda$  is not expected to alter the topology, even after (weakly) breaking spin symmetry by introducing a small Rashba SOC.

	In summary, we have shown that proximity transfers SOC properties to graphene from the TMD.\@  This effect is shown to induce different types of SOC that play an important role in determining the spin content and topological state of the system.
	Using the continuum model approach, we study the effect of lattice misalignment by either rotational or lattice mismatch or incommensuration. The resulting transferred SOC, especially the Zeeman-like type, is found to be robust and nearly insensitive to small mismatch.	
	The relatively weak intrinsic SOC, compared with the Zeeman-like SOC results likely in topologically trivial phases, as identified by the $\mathbb{Z}_2$ invariant. 

However, the effective Hamiltonian of the system may support true topologically non-trivial behavior.  We have extracted analytical condition that correctly describe gap closing and topological phases of the heterostructure. 
Reaching this fascinating regime would require the enhancement of the intrinsic SOC, perhaps via the intercalation of heavy metals and/or hydrogen decoration of graphene, in combination with large interlayer twists to help suppress the staggered potential. 
We trust that our results would encourage innovative experiments to achieve this limit.  If successful, these structures would provide the interesting possibility of achieving tunable systems in which to study topological transitions in the same physical device. 
	
	\section{Acknowledgments}
	We acknowledge support from NSF grant-DMR 1508325 (Ohio), the Saudi Arabian Cultural Mission to the US for a Graduate Scholarship. M. M. A.  acknowledges support from NSF grants DMR-1410741 and DMR-1151717. 
	
	\bibliography{cites}
	\bibliographystyle{apsrev4-1}

\end{document}